\documentclass[showpacs,amsmath,amssymb]{revtex4}
\usepackage{dcolumn}
\usepackage{bm}
\usepackage{amsthm}
\usepackage{eepic}

\newcommand{\<}{\langle}
\renewcommand{\>}{\rangle}
\newcommand{\0}{|0\rangle}
\newcommand{\1}{|1\rangle}

\newcommand{\C}{\mathbb{C}}

\newcommand{\Ha}{\mathcal{H}}
\newcommand{\E}{\mathcal{E}}
\renewcommand{\O}{\mathcal{O}}
\newcommand{\R}{\mathcal{R}}
\newcommand{\Dp}{\Delta p}
\newcommand{\id}{\operatorname{Id}}
\renewcommand{\hom}{\operatorname{Hom}}

\newcommand{\tr}{\operatorname{tr}}
\newcommand{\sign}{\operatorname{sign}}

\theoremstyle{plain} 
 
\newtheorem*{theorem*}{Theorem}
     
\newtheorem*{QEC}{Quantum Error Correction (QEC) Condition}
\newtheorem*{QEC1}{QEC condition}
\newtheorem*{QEC2}{QEC II condition}
\newtheorem*{TCF}{Thermodynamic Correlation Function (TCF) Condition}

\theoremstyle{definition}
\newtheorem{definition}{Definition}

\begin{document}

\title{Thermodynamic interpretation of the quantum error correcting
  criterion}

\author{Vladimir Korepin} \email{korepin@insti.physics.sunysb.edu}
\affiliation{Yang Institute of Theoretical Physics, State University
  of New York, Stony Brook, NY 11794-3840.}  \ \author{John Terilla}
\email{jterilla@math.sunysb.edu} \affiliation{Department of
  Mathematics, State University of New York, Stony Brook, NY
  11794-3651.}

\date{\today}

\begin{abstract}
  Shanon's fundamental coding theorems relate classical
  information theory to thermodynamics.  More recent theoretical work
  has been successful in relating quantum information theory to
  thermodynamics.  For example, Schumacher proved a 
  quantum version of Shannon's 1948 
  classical noiseless coding theorem.  In this
  note, we extend the connection between quantum information theory and
  thermodynamics to include quantum error correction.
  
  There is a standard mechanism for describing errors that may occur
  during the transmission, storage, and manipulation of quantum information.
  One can formulate a criterion of necessary and sufficient conditions
  for the errors to be detectable and correctable.  We show that this
  criterion has a thermodynamical interpretation.  
\end{abstract}


\maketitle

\section{Introduction}
Modern information theory was invented more than fifty years ago by
Claude Shannon \cite{Sh}.  In his seminal paper, he gave a
mathematical definition of information and proved his theoretical
noiseless coding and noisy coding theorems.  The noiseless theorem
describes the physical resources necessary to store and transmit the
information contained in a message.  The noisy theorem describes the
informational capacity of a noisy channel.  The basic concept behind
both theorems is that of typical messages.  As one considers messages
of increasing length, some messages emerge as likely to appear and
some messages recede and become unlikely to appear.

In statistical mechanics, one studies systems of a very large number
of particles.  The business of thermodynamics is to study the
properties of the entire system that materialize as the number of
particles grows.  Information theory as conceived by Shannon can be
compared to the physical science of one-dimensional statical
mechanics and the concept of typical sequences is like a thermodynamic
equilibrium in this statistical mechanical system.

The success of treating information theory as a physical science is by
now unquestioned and the theoretical bridge between information theory
and thermodynamics has brought a profitable trade route between the
subjects.  Presently, quantum information theory is being hotly
pursued and investigators are discovering how quantum information and
thermodynamics are connected.  One direct descendant of Shannon's
noiseless coding theorem is a quantum version proved by Schumacher
\cite{Sch2}, which shows that entropy is important in quantum
information theory.  Also, connections between information and
thermodynamics have been built using entanglement in quantum systems
\cite{RS,H1,R1}.  The bridge between information science and
statistical mechanics lengthens to include the quantum branches of
both fields and in this tradition we present a thermodynamic
interpretation of quantum error correction.

The concept of quantum error correcting codes has developed rapidly.
First quantum error correcting codes were discovered in \cite{Shor1, CSh, St1}.
Error avoiding codes were discovered in \cite{ZR}: a systematic way of
 building nontrivial models in which dynamical symmetry allows unitary
evolution of a subspace [decoherence-dissipation free] while the remaining
 part of the Hilbert space  gets strongly entangled with the environment.
Necessary and sufficient conditions for the
ability of quantum error correcting codes to correct errors appeared
\cite{BDSW, KL, CN, G, P1}.  Let us remind the reader about these
conditions.

A quantum code $C$ is a subspace of a Hilbert space $W$ equipped with
an inner product $\<\,|\,\>$.  $W$ is sometimes called an encoding
space.  Errors are represented by a collection $\E=\{E_a:W \to W\}$ of
linear operators.  One imagines the diagram $C \hookrightarrow W
\overset{\E}{\longrightarrow} W \rightarrow C $ where the middle arrow
represents deterioration of the medium while either during storage in
a quantum device, or during transmission down a noisy channel.  The
arrow on the left represents encoding the information into the
encoding space and the arrow on the right represents some kind of
recovery procedure.  The goal of quantum error correction is to
control the code space $C$, the encoding procedure, and the decoding
procedure, so that the composition of all three arrows acts as the
identity on $C$.  Here we highlight the following important result:
\begin{QEC}
  The necessary and sufficient condition for the errors $\E$ to be
  correctable is that $\<\psi | E_a^\dagger E_b |\psi \>$ be the same
  for all unit vectors $\psi \in C$ and for every $E_a, E_b \in \E$.
\end{QEC} 
In the appendix, we briefly review quantum error correction and
various critera equivalent to the QEC stated above.  In particular, we
show that the QEC criterion is equivalent to the widely known
condition that for any orthonormal basis $\{|\psi_j\>\}$ of the 
  code space $C$ and for all $E_a, E_b \in \E$, there exists a constant  
  $c_{ab}$ so that $\<\psi_j | E_a^\dagger E_b |\psi_k
\>=c_{ab}\delta_{jk}.$  Since 1996, several good quantum codes have
been developed.  Many are adapted from the classical theory of error
correction codes, the most famous being the CSS codes \cite{CSh,St1}
and their generalization--- the stabilizer codes
\cite{CRSS,CRSS2,G2,G1}.  See the textbook \cite{CN} and the report
\cite{st2} for a summary, and also the references within.

Let us now turn to statistical mechanics.  A qubit of quantum
information will be identified with a spin $\frac{1}{2}$ and we
consider a one dimensional system of interacting spins.  The state of
the system is a unit vector in a Hilbert space, which we will again
call $W$.  The dynamics of the system is determined by a Hermitian
operator $\Ha:W \to W$ called the Hamiltonian.  Akin to the emergence
of typical sequences in Shannon's theory, or the typical subspace of
Schumacher's theory, a special subspace of $W$ reveals itself as the
number of spins grows to infinity.  Let us denote this subspace, which
is called the subspace of thermodynamic equilibrium (see definition
\ref{defofC}) or just thermo-equilibrium space, by $C$.

Now, we describe some important features of the subspace $C$.  Quantum
spin chain models include a real parameter $T$ called temperature.
Usually, when $T=0$, there is a unique vector in $W$, called the
ground state, corresponding to the smallest eigenvalue of $H$.  As a
one dimensional space, the span of the ground state is too small to be
used for quantum coding.  However, for $T>0$, the space of
thermodynamic equilibrium has an exponentially large dimension (see
the line preceding equation (\ref{entropy})).  Our attitude is that
the entire space of thermo-equilibrium is a kind of `high dimensional
ground state'.  It behaves in most ways as a single state, and any
unit vector chosen from the thermo-equilibrium space will serve,
equally well, to represent the macroscopic physical properties of the
entire equilibrium.  Physical properties are properties such as
energy, scattering matrix, and most importantly for this paper, local
correlation functions.  A correlation function is a physical quantity
associated to a operator $\O:W \to W$ (see equation
(\ref{correlations}) for a definition).  It depends on $T$ and is
denoted by by $\<\O\>_T$.  We present the following fundamental
observation: For solvable models, such as XX0, XY, XYZ, Hubbard model,
etc..., one has \cite{KBI,CIKT,IIKS,IPZ}:
\begin{TCF}
  In the thermodynamic limit, the correlation functions $\<\O\>_T$ of
  local operators $\O$ satisfy the equation $\<\O\>_T=\<\psi | \O |
  \psi\>$ for any unit vector $\psi\in C$, the subspace of
  thermo-equilibrium.
\end{TCF}  

For any errors $E_a$ and $E_b$ affecting an arbitrary finite number of
qubits, the operator $\O=E_a^\dagger E_b$ will be a local operator and
the TCF conditions holds in the thermodynamic limit, thus
asymptotically delivering the quantum error correcting conditions.
 The reader may wonder how one can
manipulate encoded information, say for the purposes of
computation, if no local operators act non-trivially on the code space.
The TCF condition is strong, but does not preclude using 
thermodynamic systems as quantum processing devices since nonlocal operators
are available.  Two different approaches come immediately to mind.  One
is to use long products of Pauli operators affecting a number of qubits
comparable to the size of the entire system.  Another is to use
adiabatic changes of boundary conditions to obtain encoded gates.  
We expand the adiabatic approach in the final remarks of the conclusion.

We remark that the subspace of thermo-equilibrium is somewhat like a
positive temperature, dynamic analog of the zero temperature ground
state degeneracy that appears in topological quantum computing---a
subject that is also being investigated for its asymptotic ability to
correct errors \cite{K,FKLZ}.

\section{Shanon's noiseless coding theorem}
We briefly review Shannon's noiseless coding theorem.  Consider a
random binary variable $X$ where the probability that $X=0$ is $p$ and
the probability that $X=1$ is $1-p$.  Consider a sequence consisting
of $n$ values of $X$.  Then, the expected number of $0$'s in the
sequence is $np$ and the number of sequences with this expected number
of $0$'s is
\begin{equation}
\binom{n}{np}=\frac{n!}{(np)!(n(1-p))!}.
\end{equation}
For large $n$, we have the asymptotic result
\begin{equation}\binom{n}{np}\sim 2^{n S(p)}\text{ where } S(p)=-p
  \log p -(1-p) \log(1-p).
\end{equation}
Here, and throughout this paper, $\log$ means $\log$ base $2$.  Note
that $0 \leq S(p) \leq 1$ and $S(p)=1$ only if $p=\frac{1}{2}.$ A
sequence $b_1 b_2 \cdots b_n$ of values of $X$ that contains the
expected number of number of zeros (assuming $np$ is an integer) is
called a \emph{typical sequence}.  The probability that $b_1 b_2
\cdots b_n$ is typical approaches $1$ as $n$ approaches infinity.  So,
in order to communicate a given sequence of length $n$, one needs only
communicate which one of $2^{nS(p)}$ typical sequences is at hand.  In
this way $n$ bits can be encoded in $nS(p)$ bits.  Translated into a
coding theorem, one encodes blocks of $n$ bits by using $nS(p)$ bits
and the probability of being able to successfully decode a block
approaches one as $n$ tends to infinity.

One easily generalizes to the case that the random variable $X$ takes
values in some finite set $\{x\}$.  If $X$ takes the value $x$ with
probability $p(x)$, then one defines the Shannon entropy of $X$ to be
\begin{equation}\label{shannonentropy}
S(X)=-\sum_{x}p(x)\log(p(x))
\end{equation} and the same
analysis and conclusions hold.  The probability that a sequence $x_1
x_2 \cdots x_n$ of values of $X$ is typical approaches one as $n$
approaches infinity.

\section{Schumacher's coding theorem}
A quantum bit, or a qubit, can be represented by a vector in a two
dimensional complex linear space with an inner product.  One should
consider two vectors that differ by a nonzero scaler factor to
represent the same qubit.  An ensemble of $n$ qubits is represented by
an element of the $n$-fold tensor product of qubit spaces.  Let us fix
notation.  Let $W=\otimes_{j=1}^n W_j$, where each $W_j$ is a two
dimensional vector space with orthonormal basis $\{|0\>_j,|1\>_j\}$.
Each $W_j=\C|0\>_j\oplus\C\1_j\simeq \C^2$ is identified with a space
of qubits.  The following convenient notation is suggested for a basis
of $W$.  Every basis vector $|b_1\>_1\otimes |b_2\>_2 \otimes \cdots
\otimes |b_n\>_n$ can be referred to by the shortened name $|b_1 b_2
\cdots b_n \>$.  Bit strings of length $n$ index the basis vectors of
$W$ and, in this way, a general quantum state consisting of $n$ qubits
can be thought of as a linear combination of bit strings of length
$n$.

Let us define the density matrix associated to a state $|\psi\>\in W$.
One has the linear functional $\<\psi|$ in the dual space $W^*$ whose
value on $|\phi\>\in W$ is given by $\<\psi|\,\phi\>$.  A matrix
$\rho$ can then be identified with $|\psi\>\otimes \<\psi|$ via an
isomorphism $W \otimes W^* \simeq \hom(W,W)\simeq 2^n \times 2^n$
matrices.  It is common to drop the tensor sign and write
$\rho=|\psi\>\<\psi|$.  Of course, the information contained in $\rho$
is no different than the information contained in $|\psi\>.$ If
$|\psi\>$ is a unit vector and $\{|b\>\}$ is an orthonormal basis of
$W$, then $\rho$ has an expansion
\begin{equation}
\rho=\sum_b p(b) |b\>\<b|
\end{equation}
and the numbers $p(b)$ are nonnegative real numbers satisfying $\sum_b
p(b)=1.$ The numbers $\{p(b)\}$ define a probability distribution on
the collection $\{|b\>\}$.  One can extend the definition of $p$ to an
arbitrary $|\phi\> \in W$ by $p(|\phi\>)=\<\phi | \rho |\phi\>$.

Now, we can view the density matrix $\rho$ as a quantum random
variable which produces the state $|b\>$ with a probability of $p(b)$.
Or, one imagines a ``quantum signal source'' \cite{Sch2} which encodes
the classical bit string $b$, which is the output of a random variable
with probability $p(b)$, as the quantum state $|b\>$.  One can then
define a state $|b_1b_2\cdots b_n\>$ to be a \emph{typical state}
provided $b_1b_2\cdots b_n$ is a typical sequence \'a la Shannon.
Define the \emph{typical subspace} of $W$ to be the subspace $T$
spanned by all typical states.  Then the subspace $T$ serves to
compress $W$ as the probability that a random vector from $W$ will lie
in the subspace $T$ approaches $1$, as $n \to \infty$.  The dimension
of $T$ is $2^{nS(\rho)}$, where $S(\rho)$ is the Shannon entropy
defined in equation (\ref{shannonentropy}).  In the present context,
it is natural to note that $S(\rho)=\tr(\rho \log (\rho))$, which is
called the von Neumann entropy of $\rho$.  However, in the case that
$\{|b\>\}$ is a collection of arbitrary, not necessarily orthogonal,
states then it is the von Neumann entropy, not the Shannon entropy,
which computes the dimension of $T$.  Further refinements to the
role von Neumann entropy plays in 
quantum compression are still being developed \cite{HJW}.

\section{Quantum spin chains}
Now let us show how to formulate the quantum error correcting
condition via thermodynamics.  The subject of quantum statistical
mechanics and thermodynamics is treated in several textbooks
\cite{T,KBI}.  There are various models determined by different
Hamiltonians describing different interactions, and many, such as the
spin chain models, have relevance to quantum information theory.  For
example, programs are underway to investigate various aspects of the
XYZ family of quantum spin chain models with relevance to quantum
computing, aspects like the implementation of gates \cite{DBKBW},
decoherence free subspaces \cite{BKLW}, entanglement \cite{OAFF}, and encoded universality
\cite{BKDLW,LW1,LW2}.  Also, some two dimensional spin models arise in
topological quantum computing \cite{FKLZ,K}.

\subsection{Notation}
Let $\sigma^x, \sigma^y,$ and $\sigma^z$ be the Pauli matrices, which
act on $\C^2=\C \0\oplus\C\1$ as the matrices
\begin{equation}
\sigma^x= \begin{pmatrix}0 & 1 \\ 1 & 0 \end{pmatrix}, \quad
\sigma^y=\begin{pmatrix}0 & -i \\ i & 0 \end{pmatrix}, \quad \sigma^z=
\begin{pmatrix}1 & 0 \\ 0 & -1 \end{pmatrix}.
\end{equation}
It is common in quantum information theory to use the notation $X$,
$Y$, and $Z$ instead of $\sigma_x$, $\sigma_y$, and $\sigma_z$.  Set,
as before, $W=\otimes_{j=1}^n W_j$, where
$W_j=\C|0\>_j\oplus\C\1_j\simeq \C^2$.  Sometimes the integer $n$ is
called the length of the lattice or the size of the model, and the
reader may compare it to the length of the message in information
theory.  For any $j=1,\ldots, n$ and superscripts $\alpha=x,y,z$
define an operator $\sigma_j^\alpha: W\to W$ by
\begin{equation}
\sigma_j^\alpha\vert_{W_m}= \begin{cases}
  \id & \text{ if }j\neq m, \\
  \sigma^\alpha & \text{ if } j=m.
\end{cases}
\end{equation}
The operator $\sigma_j^\alpha$ acts non trivially on the $j$-th qubit
as the Pauli matrix $\sigma^\alpha$ and leaves all other qubits
unchanged.  One also has the matrix
\begin{equation}\label{creation}
\sigma^-:=\frac{1}{2}\left(\sigma^x-i \sigma^y\right)=
\begin{pmatrix}
  0 & 0 \\
  1 & 0
\end{pmatrix},
\end{equation}
called a creation operator.  By setting
$\sigma_j^-=\frac{1}{2}(\sigma_j^x- i\sigma_j^y)$, one can obtain
every basis vector in $W$ by acting on $\0:=|00\cdots 0\>$ by products
of the $\sigma_j^-$.

A local operator $\O:W \to W$ is defined to be a linear combination of
products of the operators $\sigma^\alpha_j$.  If $\O$ is a product of
$t$ such operators
\begin{equation}\label{wt}
\O=\sigma_{m_1}^{\alpha_1}\sigma_{m_2}^{\alpha_2}\cdots \sigma_{m_t}^{\alpha_t}
\end{equation}
with distinct $m_j$, we say that $\O$ is an operator of weight $t$.

\subsection{The partition function and thermodynamic equilibrium}
We now wish to study the model as the spins interact.  The Hilbert
space of the model is $W\simeq (\C^2)^{\otimes n}$ and the interacting
spins are governed by a Hamiltonian $\Ha:W \to W$.  What follows is
quite general, though later we illustrate more details with the XX0
model.  So the reader may have the XX0 Hamiltonian in mind:
\begin{equation} \label{ham}
\Ha = - \sum_{j=1}^n \left(
\sigma_{j}^x\sigma_{j+1}^x+\sigma_{j}^y\sigma_{j+1}^y+h\sigma_j^z\right)
\end{equation}
The real parameter $h$ is called the magnetic field.  The XX0 model
with periodic boundary conditions was originally solved in 1961 by E.
Lieb, T.  Schultz, and D. Mattis \cite{LSM}.  The XX0 model is 
sometimes called the ``isotropic XY model'' and is also known 
(in quantum information theory) as the ``XY model with Zeeman splitting.''

The thermodynamic limit of the model is mathematical idealization of a
very large system defined by a controlled limit $n\to \infty$.
Quantities of interest are computed for finite $n$ and then studied as
$n$ tends to infinity.  Often these quantities are proportional to $n$
and the proportionality factor has a finite value in the thermodynamic
limit.  These asymptotic are studied much the same way that they are
in information theory when the length of the message grows to
infinity.

The central object of thermodynamics is the partition function
$\mathcal{Z}$, which is defined at a temperature $T\geq 0$, by
\begin{equation}\label{partitionfunction}
\mathcal{Z}=\tr\left( 2^{ - \frac{\Ha}{T}}\right).
\end{equation}
In the thermodynamic limit, the partition function can be computed by
the method of steepest descent.  One has
\begin{equation}
  \mathcal{Z}=\tr 2^{\frac{-\Ha}{T}}
  = \sum_{\text{eigenvectors $v$}}2^{-\frac{E(v)}{T}}
  = \sum_{\text{eigenvalues $E$}}2^{nS} 2^{-\frac{E}{T}}
\end{equation}
where the factor $2^{nS}$ is the degeneracy of the energy level $E$
and $S$ is entropy.  Both energy and entropy increase linearly with
$n$. So evaluating
\begin{equation}\label{genlthermopartitionfunction}
\lim_{n \to \infty} \mathcal{Z}= \lim_{n \to \infty} \sum_E 2^{nS
  -\frac{E}{T}} 
\end{equation}
by the method of steepest descent leads to the variational equaiton
\begin{equation}\label{variational}
\delta\left(S - \frac{E}{nT}\right)=0.
\end{equation}
This brings us to the key definition:
\begin{definition}\label{defofC}
  The subspace of thermodynamic equilibrium is defined to be the span
  of the set of eigenvectors that solve equation (\ref{variational}).
\end{definition}
This definition of the thermo-equilibrium subspace makes sense for all
solvable models, including XX0, XXZ, XYZ, etc...  For the model we've
chosen to work with, we can be more specific.  For the XX0 model,
equation (\ref{genlthermopartitionfunction}) becomes
\begin{equation}\label{thermopartitionfunction}
\mathcal{Z}=\tr_C\left( 2^{ - \frac{\Ha}{T}}\right) \sim 2^{\frac{n}{2 \pi}
\int_{-\pi}^ \pi dp \log \left( 1 + 2^\frac{-\epsilon(p)}{T}\right)}
\end{equation}
where $\epsilon$ is given by
\begin{equation}
\epsilon(p)=-4 \cos(p)+2 h.
\end{equation}
The symbol $\tr_C$ means the trace over the subspace $C$.  
It is defined for any operator $A:W\to W$ by $\tr_C(A)=\sum_{j=1}^r a_{jj}$ 
where $a_{jj}$ are the diagonal entries of $A$ when 
expressed as a matrix using a basis 
$\{|\psi_j\>\}_{j=1}^{2^n}$ for $W$ extending a basis $\{|\psi_j\>\}_{j=1}^{r}$ for $C$.

The precise meaning of the right hand side of equation
(\ref{thermopartitionfunction}) is that $\lim_{n \to \infty}
\frac{1}{n}\log \mathcal{Z}$ exists and is given by
\begin{equation}\label{fe}
\lim_{n \to \infty} \frac{1}{n}\log \mathcal{Z}= \frac{1}{2 \pi}\int_{-\pi}^ \pi dp \log \left( 1 + 2^\frac{-\epsilon(p)}{T}\right).
\end{equation}
The quantity in equation (\ref{fe}) is called bulk free energy.

From equations (\ref{generalentropy}) and (\ref{rhosolution}), in
section \ref{thermoXX0}, we determine, asymptotically, the dimension
of $C$. The dimension of $C\sim 2^{nS}$ where $S$ is the entropy:
\begin{equation}\label{entropy}
S=- \frac{1}{2 \pi} \int_{-\pi}^\pi dp\left[ \theta(p) \log
\theta(p)
+(1-\theta(p))\log(1-\theta(p))\right],
\end{equation}
where $\theta$, called the Fermi weight, is defined by
\begin{equation}\label{fermiweight}
\theta(p)=\left(1+2^{\frac{\epsilon(p)}{T}}\right)^{-1}.
\end{equation}
Physicists call $\epsilon$ the energy of the spin wave and call $p$
the momentum of the spin wave.

\subsection{Correlation functions}
Let us define correlation functions.  Let $\O$ be a linear combination
of products of operators $\sigma_j^\alpha$.  The correlation function
$\<\O\>_T$ is defined by
\begin{equation}\label{correlations}
\<\O\>_T=
\frac{\tr \left(2^{-\frac{\Ha}{T}}\O\right)}
{\mathcal{Z}}
\end{equation}
If $\O$ has weight $t$, then $\<\O\>_T$ can be related to $t$-point
correlation function.  In the thermodynamic limit, the correlation
functions also can be calculated explicitly by the method of steepest
descent, just as for the partition function
$\mathcal{Z}=\tr(2^{-\frac{\Ha}{T}})$.  Only the space of thermo
equilibrium $C$ contributes to the trace (\cite{KBI}, page 25).  So
instead of taking the trace over all of $W$, one has
\begin{equation}\label{thermocor}
\<\O\>_T=\frac{\tr_C \left(2^{-\frac{\Ha}{T}}\O\right)}
{\mathcal{Z}}.
\end{equation}
An even stronger statement is true. In \cite{KBI,CIKT,IIKS,IPZ} it was
shown that each term of the trace in equation (\ref{thermocor})
contributes equally, and so equation (\ref{thermocor}) simplifies
further:
\begin{equation}\label{thermocor2}
\<\O\>_T=\frac{\tr_C
  \left(2^{-\frac{\Ha}{T}}\O\right)}{\mathcal{Z}}=\frac{\<\psi|\O|\psi\>}{\<\psi|\psi\>}
  \text{ for any }\psi\in C,
\end{equation}
which gives the thermodynamic correlation function condition stated in
the introduction.

The TCF condition is a quite general feature of quantum statistical
mechanics.  It holds not just for XX0, but also for other integrable
models (XY, XXZ, XYZ, nonlinear Schrodinger, Hubbard model, etc...).
We conjecture that the TCF condition is valid for a wide class of
physically interesting models, including non-integrable models in some
vicinity of an integrable one.

Let us comment on the the special case $T=0$. The correlation
functions simplify dramatically.  There is a unique vector $|G\>$,
first identified in \cite{LSM}, called the ground state.  It
corresponds to the lowest eigenvalue of $\Ha$.  In the case of zero
temperature
\begin{equation}\label{zerotemp}
\<\O\>_{T=0}=\<G|\O|G\>.
\end{equation}
The TCF condition is precisely a generalization the equation
(\ref{zerotemp}) to $T>0$.

\subsection{Eigenvectors of the XX0 Hamiltonian}

Except for equations
(\ref{thermopartitionfunction}-\ref{fermiweight}), the discussion
above applies to a quantum statistical model governed by most any
Hamiltonian.  Now, to further illustrate the thermodynamics, we work
specifically with XX0.  The Hamiltonian $$\Ha = - \sum_{j=1}^n
\left(\sigma_{j}^x\sigma_{j+1}^x+\sigma_{j}^y\sigma_{j+1}^y+h\sigma_j^z\right)$$
can be written as $\Ha=\Ha_0- 2 h S^z$ where
\begin{equation} 
\Ha_0 = - \sum_{j=1}^n\left(
\sigma_{j}^x\sigma_{j+1}^x+\sigma_{j}^y\sigma_{j+1}^y \right)\text{ and
  }S^z=\frac{1}{2}\sum_{j=1}^n \sigma_j^z.
\end{equation}
Note that $[\Ha_0, S^z]=0$.  The problem of finding the eigenvectors
of $\Ha$ can be reduced to finding the common eigenvectors of $\Ha_0$ and
$S^z$.  We now describe a complete set of eigenvectors of $\Ha$ for any positive integer $n$.  
  Recall that
$\sigma_j^-=\frac{1}{2}(\sigma_j^x- i\sigma_j^y)$.

The eigenvectors of $\Ha$ are determined by a positive integer $m\leq
n$ and a collection of real numbers $\{p_i\}_{i=1}^m$, with each
$-\pi<p_i<\pi$, called momenta.  For short, the collection
$\{p_i\}_{i=1}^m$ may be denoted simply by $\{p\}$.  Define a vector
$|\{p\}\>_m\in W$ by
\begin{equation}\label{eigenvectors}
|\{p\}\>_m=
\frac{1}{\sqrt{m!}}\sum_{x_1, \ldots, x_m} 
\chi_m(\{x\}|\{p\})
\sigma_{x_m}^- \cdots \sigma_{x_1}^- \0.
\end{equation}
The complex valued function $\chi$ is defined by
\begin{equation}\label{wavefunctions}
\chi_m\left(\{x\}|\{p\}\right)=
\frac{1}{\sqrt{m!}} \left( \prod_{1 \leq
    a < b \leq m} \sign(x_b-x_a)\right) \det (A),
\end{equation}
where $A$ is the $m\times m$ matrix with $(j,k)$ entry $A_{jk}=\exp(i
x_j p_k)$ and the sign function is given by

\begin{equation}
\sign(x)=
\begin{cases}
  1 & \text{ if }x>0,\\
  -1 & \text{ if }x<0,\\
  0 & \text{ if }x=0.
\end{cases}
\end{equation}  
The function $\chi(\{x\}|\{p\})$ is called a wave function.  It is
symmetric in $x$ and antisymmetric in $p$, so we assume that $x_1\leq
x_2 \leq \cdots \leq x_m$ and that $p_1< p_2 < \cdots < p_n$.  We may
drop the subscript $m$ from $|\{p\}\>_m\in W$ if the size of $\{p\}$
does not need emphasis and just write $|\{p\}\>$.

The periodic boundary conditions of the model lead to the following
equation for each $p_j$:
\begin{equation}\label{momentaeqns}
\exp(i p_j n)=(-1)^{m+1}
\end{equation}
Note that (\ref{momentaeqns}) has $n$ solutions and a collection
$\{p\}$ amounts to a choice of $m$ of these solutions.  So, for each
$m=0, \ldots, n$, there are $\binom{n}{m}$ choices of momenta each
identifying one eigenvector of $\Ha$.  In total, there are $2^n$
eigenvectors (\ref{eigenvectors}) as $m$ varies from $0$ to $n$.  A
straightforward calculation of their scalar products shows that these
eigenvectors form an orthogonal basis of $W$:
\begin{equation*}
  \<\{p\}|\{p\}\>=n^m \text{ and }\<\{p\}|\{q\}\>=0 \text{ if }\{p\}\neq \{q\}.
\end{equation*}
The eigenvalue $E(\{p\})$ of $|\{p\}\>_m$ is given by
\begin{equation}
E(\{p\})=\sum_{j=1}^m \epsilon (p_j),  \quad \epsilon(p)=-4 \cos p + 2h.
\end{equation}

\subsection{Thermodynamics of the XX0 model}\label{thermoXX0}
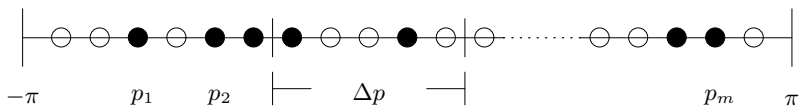
\begin{figure}[h]
\caption{A picture of typical positions and momenta}\label{particlefig}
\setlength{\unitlength}{0.001in} {
\begin{picture}(5000,600)(-500,100)
\put(200,400){\ellipse{100}{100}}
\put(400,400){\ellipse{100}{100}}
\put(600,400){\blacken\ellipse{100}{100}}
\put(800,400){\ellipse{100}{100}}
\put(1000,400){\blacken\ellipse{100}{100}}
\put(1200,400){\blacken\ellipse{100}{100}}
\put(1400,400){\blacken\ellipse{100}{100}}
\put(1600,400){\ellipse{100}{100}}
\put(1800,400){\ellipse{100}{100}}
\put(2000,400){\blacken\ellipse{100}{100}}
\put(2200,400){\ellipse{100}{100}}
\put(2400,400){\ellipse{100}{100}}
\put(3000,400){\ellipse{100}{100}}
\put(3200,400){\ellipse{100}{100}}
\put(3400,400){\blacken\ellipse{100}{100}}
\put(3600,400){\blacken\ellipse{100}{100}}
\put(3800,400){\ellipse{100}{100}}
\path(0,550)(0,250)
\path(4000,550)(4000,250)
\path(1300,500)(1300,300)
\path(2300,500)(2300,300)
\path(1300,220)(1300,50)
\path(2300,220)(2300,50)
\path(1300,135)(1500,135)
\path(2100,135)(2300,135)
\path(0,400)(2500,400)
\path(2900,400)(4000,400)
\dottedline{45}(2500,400)(2900,400)
\put(0,50){\makebox(0,0)[cb]{$-\pi$}}
\put(625,50){\makebox(0,0)[cb]{$p_1$}}
\put(1025,50){\makebox(0,0)[cb]{$p_2$}}
\put(1800,50){\makebox(0,0)[cb]{$\Dp$}}
\put(3625,50){\makebox(0,0)[cb]{$p_m$}}
\put(4000,50){\makebox(0,0)[cb]{$\pi$}}
\end{picture}
}
\end{figure}

One may think of each of the eigenvectors of $\Ha$ as being obtained
from the ferromagnetic state $\0$ by adding $m$ particles (flipping
$m$ spins) via the creation operators $\sigma^-_j$, with momenta
$\{p_j\}$ and energies $\{\epsilon(p_j)\}$.  In this section we
consider the situation as the number of particles occupying positions
in the interval $[-\pi,\pi]$ tends to infinity (see \cite{KBI}).
Divide the interval $[-\pi,\pi]$ into $n$ subintervals, each of length
$\frac{2 \pi}{n}.$ Each subdivision point will be called a position
and should be thought of as a possible location for a particle.  They
represent the $n$ solutions of (\ref{momentaeqns}), which the momenta
satisfy.  When a wave function has been selected, and a choice $\{p\}$
of momenta has been made, one should think that each of the $m$
positions appearing in $\{p\}$ is occupied by a particle and the
remaining $n-m$ positions are empty (see figure (\ref{particlefig})).

Once a collection $\{p\}=\{p_1 < p_2 < \cdots < p_n\}$ has been
chosen, one may look at the numbers $\rho(p_j)$:
\begin{equation}\label{rhoparticles}
\rho(p_j)=\frac{1}{n(p_{j+1}-p_j)}.
\end{equation}
Now we consider the situation where the number of positions becomes
large, the number of particles becomes large, and the density of
particles remains fixed.  The thermodynamic limit is defined by
\begin{equation}
m\to \infty, \quad n \to \infty, \quad \text{and } d:=\frac{m}{n} \text{ is
  constant}.
\end{equation}
The reader may compare it to the double-scaling limit in matrix
models.  For now, consider $m$ and $n$ as very large, but still
finite.  As the number of particles grows, it becomes difficult to
keep track of the individual momenta $p_j$ since there are so many of
them.  Instead, the growing collections $\{p\}$ give way to a finite
function $\rho:[-\pi,\pi]\to \mathbb{R}$, which exists as a limit of
expressions (\ref{rhoparticles}).  It has the interpretation that for
large $n$
\begin{equation*}
\text{ the number of particles in the interval}[p,p+\Dp] \approx n \rho(p)\Dp,
\end{equation*}
provided the scale for $\Dp$ is chosen properly.  One should have
\begin{equation}\label{scale}
\frac{2\pi}{n}<<\Dp,
\end{equation}
but $\Dp$ should still be small enough to be able to approximate
$\rho$ by a constant on the interval $[p,p+\Dp]$.

One should think that $\rho$ is part of a macroscopic description of
the model, and that a collection $\{p\}$ is part of a microscopic
description.  Be aware that each $\rho$ may be the limit of many
different sequences of microscopic states $\{p\}$, the number of which
can be called the degeneracy of the macroscopic state $\rho$.  For a
fixed $\rho$, we can compute the number of microscopic states
corresponding to $\rho$.  First, partition the interval $[-\pi,\pi]$
into subintervals of length $\Dp$.  It is important that the scale of
$\Dp$ is such that $\rho$ is well approximated by a constant on the
interval $[p,\Dp]$.  The number of ways of selecting $n \rho(p)\Dp$
positions from $\frac{n}{2 \pi} \Dp$ vacancies in which to place
particles is equal to the binomial coefficient:

\begin{equation}\label{combinations}
\binom{\frac{n}{2 \pi} \Dp}{n \rho(p)\Dp}=\frac{\left(\frac{n}{2 
\pi}\Dp\right) !}{\left(n\rho(p) \Dp\right ) ! \; \left 
(n\left(\frac{1}{2\pi}-\rho(p)\right) \Dp\right ) !}.
\end{equation}

Using Stirling's formula to approximate a factorial, one arrives
at equation (\ref{combinations}) asymptotically becomes $2^{n \Delta
  S}$ where
\begin{equation}
\Delta S=\left[ \frac{1}{2 \pi}\log \left( \frac{1}{2 \pi}\right)-
  \rho(p)\log(\rho(p))- \left(\frac{1}{2\pi}-\rho(p)\right)\log\left(\frac{1}{2\pi}-\rho(p)\right)\right]\Dp.
\end{equation}
Now, as $\Dp \to 0$, we find that asymptotically, $\rho$ has a
degeneracy of $2^{nS}$ where $S$, called the entropy, has the form
\begin{equation}\label{generalentropy}
S= \int_{-\pi}^\pi dp \left[ \frac{1}{2 \pi}\log \left( \frac{1}{2 \pi}\right)-
  \rho(p)\log(\rho(p))-
\left(\frac{1}{2\pi}-\rho(p)\right)\log\left(\frac{1}{2\pi}-\rho(p)\right)\right] 
\end{equation}

The following picture emerges.  For a finite system, each collection
$\{p\}$ corresponds to precisely one $|\{p\}\>\in W$, which is an
eigenvector of $\Ha.$ In the thermodynamic limit when the number of
spins becomes infinite, we replace the momenta $\{p\}$ by the function
$\rho$.  But unlike the finite system, each $\rho$ corresponds to many
eigenvectors in the now infinite dimensional Hilbert space---the
different microscopic states corresponding to a single macroscopic
state.  Moreover, each of these eigenvectors have the same energy
density (see equation (\ref{rhoenergy}) below).  The fact that every
eigenvector corresponding to a single $\rho$ has the same energy
density is an instance of a crucial principle.  Namely, \emph{all of
  the local observables in the thermodynamic limit depend only on the
  macroscopic variable $\rho$.}  Consider, for instance, the energy
$E(\{p\})$:
\begin{equation}
E(\{p\})=\sum_{j=1}^m \epsilon (p_j)=n \sum_{j=1}^m \epsilon
(p_j)\frac{p_{j+1}-p_j}{n(p_{j+1}-p_j)}.
\end{equation} 
In the thermodynamic limit, one finds that energy density is equal to
\begin{equation}\label{rhoenergy}
\lim_{n \to \infty}\frac{E}{n}= \int_{-\pi}^{\pi} \epsilon(p)\rho(p) dp.
\end{equation}
The thermodynamic correlation function condition is a consequence of
the principle stated above.  The matrix elements
\begin{equation}
\lim_{n \to \infty}\frac{\<\{p\}|\O|\{p\}\>}{\<\{p\}|\{p\}\>}
\end{equation}
depend only on $\rho$, not on the set $\{p\}$.  Therefore, in the
thermodynamic limit,
\begin{equation}
\<\O\>_T=\frac{\tr\left(2^{-\frac{\Ha}{T}}\O\right)}{\mathcal{Z}}=\lim_{n \to \infty}\frac{\<\{p\}|\O|\{p\}\>}{\<\{p\}|\{p\}\>}
\end{equation}

In order to determine which function $\rho (p)$ defines the space of
thermo equilibrium, we return to the variational equation, which arose
from the steepest descent approximation.  Using variational calculus,
one has for XX0
\begin{gather}\label{ds}
  \delta S = \int_{-\pi}^{\pi} dp \left[-\delta \rho(p) \log (\rho
    (p))+ \delta \rho(p) \log \left
      (\frac{1}{2\pi}-\rho(p)\right)\right] \intertext{and}
\label{de}
\delta E =\int_{-\pi}^{\pi} dp \left[-\delta \rho(p)
  \epsilon(p)\right].
\end{gather}
Setting $\delta\left(S - \frac{E}{nT}\right)=0$ gives the solution
\begin{equation}\label{rhosolution}
\rho(p) =\frac{1}{2 \pi}\theta(p)=\frac{1}{2 \pi
  \left(1+2^{\frac{\epsilon(p)}{T}}\right)}.
\end{equation}  
Therefore, the span of the eigenvectors that correspond to the
function $\rho=\frac{1}{2 \pi} \theta$ comprise the space $C$ of
thermodynamic equilibrium.  Since every eigenvector in the
thermo-equilibrium space corresponds to the same $\rho$, the
correlation functions $\<\psi |\O|\psi \>$ do not depend on which
$\psi\in C$ is chosen.  It can be a challenge to compute these
correlation functions explicitly.  In \cite{CIKT, KBI,IIKS}, they are
computed for several models including XX0 by direct calculations.

\section{Concluding remarks}
In statistical mechanics, the probability that an eigenstate
$|\psi\>\in W$ will appear is given by
$\frac{1}{\mathcal{Z}}2^{-\frac{E(|\psi\>)}{T}}$ where the energy
$E(|\psi\>)$ is the eigenvalue of the eigenvector $|\psi\>$.  This
leads to a definition of a typical state and the subspace of
thermo-equilibrium becomes comparable to Schumacher's typical
subspace.  We have shown that this subspace of thermo-equilibrium
satisfies, asymptotically, the quantum error correction criterion for
all errors of finite weight, thus strengthening the bond between
information science and statistical mechanics.  The reader may imagine
that for finite $n$, an approximation of the thermo-equilibrium space
may serve as quantum code.  In order to develop this idea, one needs
ways of producing gates and making measurements.  Gates in quantum
spin chain models have already been implemented in several
circumstances \cite{KW, TD, BDS}.  Presumably, gates operating in the
thermo-equilibrium subspace (or some finite dimensional approximation
of it) can be obtained by similar means.  Short products of Pauli
matrices cannot be employed to map states in the thermo-equilibrium
space into one another since such products are local operators and
represent exactly the errors from which the thermo-equilibrium space
protects against.  However, products of Pauli matrices having a number of
factors that is proportional to the length of the lattice $n$ (the
length of the quantum message) are candidates for gates.

As we mention in the introduction, another promising approach for
developing gates in thermodynamic codes is more topological.  We wrote
this paper assuming periodic boundary conditions.  That is, the wave
function $\chi_m(\{x\} |\{p\})$ satisfies
\begin{equation}
\chi_m(x_1+n, x_2, \ldots , x_n |\{p\})=\chi_m(x_1,x_2  \ldots x_n |\{p\}).
\end{equation}
Thus, one can imagine the lattice forming a circle.  Now, by
introducing a magnetic flux threading this circle, the boundary
conditions become twisted by a real phase $\phi$:
\begin{equation}
\chi_m(x_1+n, x_2, \ldots , x_n |\{p\})= e^{i\phi}
\chi_m(x_1,x_2  \ldots x_n |\{p\}).
\end{equation}
By adiabatically changing $\phi$ from $0$ to $2 \pi$ the
subspace of thermodynamic equilibrium, i.e. the codes space, will be
mapped into
itself.  Even at zero temperature, this map is nontrivial---Berry's phase for such an adiabatic process was calculated in \cite{KWu}.

\appendix

\section{Quantum error correction criteria}

The occurrence of errors during storage or transmission of quantum data
is governed by a quantum operation, also called a super-operator.
Given a finite set of linear transformations $\E=\{E_a:W \to
W\}_{a=1}^m$ satisfying $\sum_{a=1}^m E_a^\dagger E_a=\id$, one
defines a super-operator $S_\E$ acting on density matrices describing
states in $W$.  The action of $S_\E$ is defined on a density matrix
$P$ describing states in $W$ by
\begin{equation}
S_\E( P )=\sum_{a=1}^m E_a P E_a^\dagger.
\end{equation}
Let $C\subset W$ be a quantum code.  One says that the code $C$ can
correct the errors $\E$, or that the errors $\E$ are correctable,
provided there exists another super-operator $S_\R$ (expressed by a
collection $\R=\{R_b\}_{b=1}^s$ satisfying $\sum_{b=1}^s R_b^\dagger
R_b=\id$) such that
\begin{equation}
S_\R(S_\E(P))=P \text{ for all density matrices $P$ describing states
  in $C\subset W$.}
\end{equation}
We now recall (theorem 10.1 from \cite{CN})
\begin{theorem*}Let 
  $\pi:W \to C$ be the orthogonal projector onto the code subspace and
  $\E=\{E_a:W \to W\}_{a=1}^m$ be a collection of linear operators
  with $\sum_{a=1}^m E_a^\dagger E_a=\id$.  A recovery super-operator
  $S_\R$ inverting $S_\E$ on density matrices from $C$ exists if and
  only if for every $E_a,E_b \in \E$, there exists a constant $c_{ab}$
  satisfying $\pi E_a^\dagger E_b \pi = c_{ab} \pi$.
\end{theorem*}
This theorem is evidently equivalent to
\begin{QEC2} 
  In order for the errors $\E$ to be correctable, it is necessary and
  sufficient that for any orthonormal basis $\{|\psi_j\>\}_{j=1}^l$ of
  the code space $C$ and for each $E_a, E_b \in \E$, there exists a
  constant $c_{ab}$ so that $\<\psi_j | E_a^\dagger E_b |\psi_k
  \>=c_{ab}\delta_{jk}$.
\end{QEC2}
The condition QEC II as stated above seems to be familiar to those
working in the field and can be found in many places (for example,
section 3 of \cite{KL}, in chapter 7 (pages 9 and 86) of \cite{P3},
and section 6.4 of \cite{BDSW}.)  In the introduction, we stated the
\begin{QEC1}
  The necessary and sufficient condition for the errors $\E$ to be
  correctable is that $\<\psi | E_a^\dagger E_b |\psi \>$ be the same
  for all unit vectors $\psi \in C$ and for every $E_a, E_b \in \E$.
\end{QEC1}
We now prove that the conditions QEC and QEC II are equivalent.

\begin{proof}Suppose that
  $\<\psi | E_a^\dagger E_b | \psi\>=c_{ab}$ for every unit vector
  $|\psi\> \in C$.  Note that the matrix $A=[c_{ab}]$ is Hermitian,
  hence there exists a unitary $U=[u_{rs}]$ such that
  $UAU^{\dagger}=D=[d_{ab}]$ is a real diagonal matrix.  Now let us
  define $F_r=\sum_{a=1}^m u_{ra}E_a.$ Note that for any unit vector
  $\psi \in C$, we have
\begin{equation}\label{eigenvalue}
\<\psi | F_r^\dagger F_s | \psi \>= d_{rr}\delta_{rs}.
\end{equation}
The operator $F_r^\dagger F_r$ is Hermitian so we can find an
orthonormal basis $\{|\phi_j\>\}_{j=1}^l$ for $C$ of eigenvectors of
$F_r^\dagger F_r$.  By substituting $|\phi_j\>$ for $|\psi\>$ in
equation \ref{eigenvalue}, one finds that the eigenvalue corresponding
to $|\phi_j\>$ is $d_{rr}$ for every $j$.  Thus,
$F^\dagger_rF_r=d_{rr}I.$ It follows that for any orthonormal basis
$\{|\psi_j\>\}_{j=1}^l$ of $C$
\begin{equation}
\<\psi_j | F_r^\dagger F_s | \psi_k \>= d_{rr}\delta_{rs}\delta_{jk}.
\end{equation}
By changing back from the $\{F_r\}$ to the $\{E_a\}$ by using
$E_a=\sum_{r=1}^m \bar{u}_{ra}F_r$, we find that
\begin{equation}
\<\psi_j | E_a^\dagger E_b | \psi_k \>= c_{ab}\delta_{jk}.
\end{equation}
This proves that $\mathrm{QEC}\Rightarrow\mathrm{QEC II}.$

Now, suppose that $\{|\psi_j\>\}_{j=1}^l$ be an orthonormal basis for
$C$ and that $\<\psi_j | E_a^\dagger E_b | \psi_k \>=
c_{ab}\delta_{jk}.$ For any unit vector $\psi\in C$ we have
$\psi=\sum_{j=1}^l a_j \psi_j$ for some $a_j$ with $\sum_{j=1}^l 
\bar{a}_ja_j=1$.
We compute
\begin{align}
  \<\psi | E^\dagger_aE_b | \psi\>&=\sum_{j=1}^l\sum_{k=1}^l \bar{a}_j
  a_k \<\psi_j
  | E^\dagger_aE_b  | \psi_k\>\\
  &=\sum_{j=1}^l \bar{a}_ja_j c_{ab}\\
  &=c_{ab}.
\end{align}
Thus $\mathrm{QEC II}\Rightarrow\mathrm{QEC}.$ In such a way we proved
the equivalence of QEC II and QEC.
\end{proof}

\end{document}